\documentstyle[graphicx,twocolumn,aps]{revtex}

\tightenlines

\begin{document}

\newcommand{\bq}{\begin{equation}}
\newcommand{\eq}{\end{equation}}
\newcommand{\etal}{{\it et al.}}

\draft

\preprint{ND Atomic Theory Preprint 06/99}

\title{Many-body and model-potential calculations of low-energy 
        photoionization parameters for francium}

\author{A. Derevianko and  W. R. Johnson}
\address{Department of Physics, Notre Dame University, Notre Dame, IN 46556}
\author{H. R. Sadeghpour}
\address{Institute for Theoretical Atomic and Molecular Physics\\
         Harvard-Smithsonian Center for Astrophysics, Cambridge, MA 02138}

\date{\today}

\maketitle

\begin{abstract}
The photoionization cross section $\sigma$, spin-polarization parameters 
$P$ and $Q$, and the angular-distribution asymmetry parameter $\beta$ are calculated 
for the $7s$ state of francium for photon energies below 10 eV. 
Two distinct calculations are presented, one based on many-body perturbation 
theory and another based on the model potential method. Although predictions of the
two calculations are similar, the detailed energy dependence of the 
photoionization parameters from the two calculations differ.
From the theoretical $p$-wave phase shifts, we infer quantum defects for  
$p_{1/2}$ and $p_{3/2}$  Rydberg series, permitting us to calculate positions 
of experimentally unknown $p$ states in francium.
\end{abstract}

\pacs{31.15.Md, 32.80.Fb, 33.60.-q}

\section{Introduction}

Remarkable progress has been made recently in determining 
energies and lifetimes of low-lying states of 
the heaviest alkali-metal atom francium\cite{exf}, 
motivated in part by the enhancement of parity non-conserving (PNC) effects in francium
compared with other  alkali-metal atoms.   
This experimental work has been accompanied by theoretical studies of
properties of the francium atom \cite{thf,Marinescu98}, 
concerned mostly with energies and hyperfine constants of the
ground and low-lying excited states or transitions between such states.

In this work, we present two calculations of photoionization of francium
for photon energies below 10 eV; the first is an {\em ab-initio} many-body calculation 
and the second is a model potential (MP) calculation. Experiments on photoionization of
francium are planned for the Advanced Light Source at Berkeley.\cite{HG}

{\em Ab-initio} calculations of photoionization
in alkali-metal atoms have proved to be a formidable challenge.  
Photoionization calculations  in cesium based on the Dirac
or Breit-Pauli equations \cite{CK,OM,HS} accounted for  
the spin-orbit interaction, but not for shielding of the dipole
operator by the core electrons or for core-polarization effects; 
whereas, relativistic calculations that included corrections from 
many-body perturbation theory (MBPT) at the level of the 
random-phase-approximation (RPA) \cite{JS,FJ} accounted 
for the spin-orbit effects and for core shielding, but not core polarization.
Predictions from these many-body calculations were in poor agreement with 
measurements of the Fano
spin-polarization parameter $P$ by Heinzmann {\it et al.}\ \cite{Fano69}, 
of the spin-polarization parameter
$Q$ by Lubell {\it et al.}\ \cite{LubellRaith69}, 
and with the measurement of the angular-distribution asymmetry
parameter $\beta$ by Yin and Elliott \cite{Yin}.  
The first quantitatively successful many-body calculation of photoionization of 
cesium was a relativistic many-body calculation
that included both core polarization and core shielding corrections \cite{WRJ_Indian97};
that method is applied to low-energy photoionization of francium in the present paper.

Although successful many-body calculations of photoionization of heavy alkali-metal atoms
are of recent vintage, 
nearly three decades ago, a number of increasingly 
sophisticated and successful model potential calculations of the photoionization 
of cesium appeared \cite{Hameed,Beigman,WD,Weisheit}, culminating 
with that of Norcross \cite{Norcross}.  
The latter calculation, which included the spin-orbit interaction, long-range polarization 
potentials, and shielding corrections to the dipole operator, gave
quantitatively correct values for all of the measured photoionization parameters
in cesium.  A model potential similar to the one used in \cite{Norcross} was developed 
recently \cite{Marinescu98}  
to study transitions in francium and is used here to study 
low-energy photoionization in francium. 

Below, we sketch the important features of the theoretical methods. The photoionization cross
sections for Fr $7s$ are calculated in Section II.B in both methods and are compared against 
one another. In Section II.E, we give results for the spin-polarization parameters and the 
angular distribution asymmetry parameter. Section III concludes our discussion of the francium 
photoionization.

\section{Theoretical Analysis \& Discussion}
     
\subsection{Many-Body Perturbation Theory}

\begin{figure}
\centerline{\includegraphics[scale=0.65]{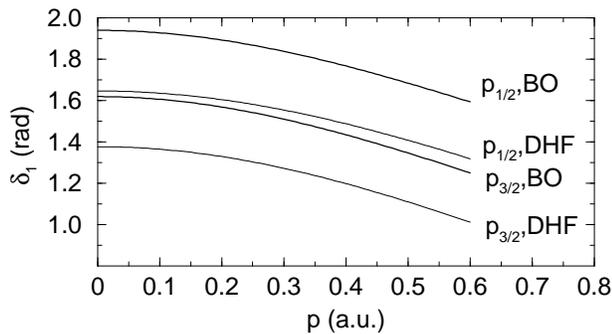} }
\caption{Short-range phase shifts for $p$ continuum in Fr calculated
  in Dirac-Hartree-Fock (DHF) approximation and including
  self-energy correction (BO). \label{Fig-phases}} 
\end{figure}

We start our many-body analysis from the Dirac-Hartree-Fock (DHF) 
$V_{N-1}$ approximation, in which the DHF equations are solved 
self-consistently for core orbitals, and the valence orbitals 
are determined subsequently in the field of the ``frozen core''.
The total phase shift $\bar{\delta}_\kappa$ for a continuum state with
angular quantum number $\kappa$ 
in the field of the core is
a sum of rapidly varying Coulomb phase shift $\delta^{\rm C}_\kappa$ and 
the short-range shift $\delta_\kappa$. 
The short-range DHF phase-shifts for $p_{1/2}$ and
$p_{3/2}$ continuum  wave functions are shown in Fig.~\ref{Fig-phases}.
The $p_{3/2}$ wave function lags in phase compared to the $p_{1/2}$
wave function owing to the spin-orbit interaction, which is attractive for
$p_{1/2}$ states and repulsive for $p_{3/2}$ states.  The DHF approximation
typically underestimates removal energies of bound electrons in heavy atoms 
such as francium by about 10\%; similar accuracy is expected for phase shifts.
To improve this level of accuracy one must take into account 
higher-order MBPT corrections.

The clear advantage of the $V_{N-1}$ approximation stems from the fact that
one-body contributions to  the 
residual Coulomb interaction vanish. This leads to a
significant reduction in the number of terms in the order-by-order
MBPT expansion. In particular, first-order corrections
to the energy (or the phase shift) vanish and 
the perturbation expansion starts in second-order. 

\begin{figure}
\centerline{\quad\includegraphics[scale=0.65]{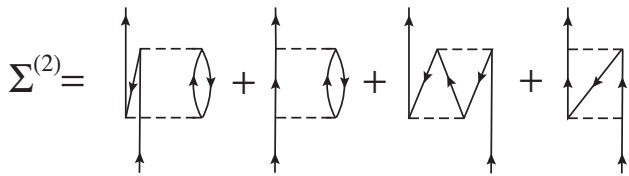} }
\caption{ The second-order self-energy operator. \label{fig-sigma}} 
\end{figure}

The leading correlation contribution to the energy 
is the expectation value of the second-order self-energy operator $\Sigma^{(2)}$, 
given diagrammatically by the Brueckner-Goldstone diagrams of Fig.~\ref{fig-sigma}. 
Solutions to the
Dirac equation including the $V_{N-1}$ potential and the self-energy operator
are called Brueckner orbitals (BO).
The non-local self-energy operator $\Sigma$, in the limit of large $r$,
describes the interaction of an electron with the induced electric 
moments of the core,
\begin{equation}
 \Sigma(r,r',\epsilon) \rightarrow 
   -\frac{\alpha_{\rm d}}{2 r^4} \, \delta(r-r') \,,
\end{equation}
where $\alpha_{\rm d}$ is the dipole polarizability of the core.
We determine second-order correction to the phase shift perturbatively as
\begin{equation}
  \delta_\kappa^{(2)} = - \sin^{-1} ( \pi \ 
\langle u_{\epsilon \kappa} | \Sigma^{(2)} | u_{\epsilon \kappa} \rangle) \, .
\label{phase}
\end{equation}
Here, $u_{\epsilon \kappa}$ is a continuum DHF wave function normalized on the
energy scale. The resulting DHF+BO phase shifts are presented in Fig.~\ref{Fig-phases}. 
The attractive polarization potential draws in the nodes of the wave function,
resulting in larger phase shifts. The change in the  
phase shift is approximately the same for both $p_{3/2}$ and $p_{1/2}$ continuum
states, demonstrating that the self-energy correction is mainly due to 
the accumulation of phase outside of the core. 

\subsection{Model Potential}

The parametric model potential used in this work has the form \cite{msd}
\bq
V_\ell^{(j)}(r)= \frac{Z_{\ell j}(r)}{r}-\frac{\alpha_d}{2r^4}[1-e^{-(r/r_c^{(j)})^6}],
\label{Eq:potential} 
\eq
where $\alpha_d$ is the static dipole polarizability of the Fr$^+$ ionic core and the effective radial
charge $Z_{\ell j}(r)$ is given by
\bq
Z_{\ell j}(r) = 1+(z-1)e^{-a_1^{(j)}r}+r(a_3^{(j)}+a_4^{(j)}r)e^{-a_2^{(j)}r}.
\label{Eq:EffZ}
\eq
The angular momentum-dependent parameters, $a_i^{(j)}, \ i=1,...,4$ and the cut-off radius $r_c^{(j)}$ are obtained
through a non-linear fit to one-electron Rydberg energy levels in francium \cite{exf,thf,expt}. 
Because the spin-orbit effects are 
appreciable for heavy alkali metals, two separate nonlinear fits; one for each fine-structure
series, $j_+=\ell +\frac{1}{2}$
and $j_-=\ell -\frac{1}{2}$ were performed. 
The static dipole polarizability was obtained from an extrapolation of the known core polarizabilities for the other alkali metals as $\alpha_d(0)=23.2$ a.u. \cite{Marinescu98}. We note that an {\it ab initio}
value for the francium core polarizability is now available\cite{thf}. 

\begin{figure}
\centerline{\includegraphics[scale=0.65]{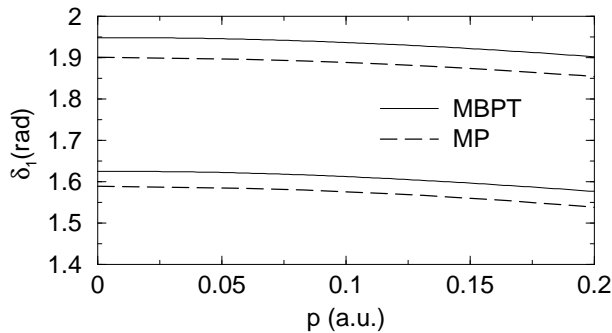} }
  \caption{Comparison of short-range phase shifts obtained in model potential
and many-body methods. Upper two curves represent $p_{1/2}$-phase
shifts, and two lower curves represent those for $p_{3/2}$ continuum.
\label{Fig-phasesCmp}} 
\end{figure}

A comparison of short-range phase shifts calculated in the model-potential method
and the MBPT is presented in Fig.~\ref{Fig-phasesCmp}. We find  reasonable agreement
between the two methods. The MP continuum wavefunctions are slightly lagging in phase
compared to many-body wavefunctions. Such phase differences result in
Cooper minima being shifted to higher photoelectron momentum in
the model-potential calculation.

\subsection{Quantum defects}
 In quantum defect (QD) theory~\cite{Seaton58}, 
the energy levels of the valence electron are 
described by a hydrogen-like Rydberg-Ritz formula 
\begin{equation}
\epsilon_{n\kappa} = -\frac{1}{2(n-\mu_\kappa)^2} 
\end{equation} 
in terms of a  quantum defect $\mu_\kappa$, 
which is represented as an expansion in powers of energy
with constant coefficients $\mu^{(i)}_\kappa$
\begin{equation}
\mu_\kappa = \mu^{(0)}_\kappa + \mu^{(1)}_\kappa \epsilon_{n\kappa} + 
\mu^{(2)}_\kappa (\epsilon_{n\kappa})^2 + \cdots \, .
\end{equation}
The Rydberg-Ritz formula provides an accurate fitting expression for the bound
spectrum of alkalis. The QD $\mu^{(0)}_\kappa$ is related to threshold value
of the phase shift as $\mu^{(0)}_\kappa = \delta_\kappa/\pi + {\rm integer}$.
The QD's for Fr $p$ states are not known, since the relevant 
Rydberg series have not been observed experimentally. We use our
{\em ab-initio} threshold phase shifts together with experimentally
known energies for $7p$ and $8p$ states to predict QD's;
thereby approximating the entire Rydberg spectrum of Fr $p$ states.
The predicted quantum defects are given in Table~\ref{Tab_mu}. We assigned
an error bar of $0.5\%$ to the threshold phase shift, based on the accuracy
of an application of the many-body
formalism employed here to the case of Cs~\cite{WRJ_Indian97}. 

In Table~\ref{Tab_mu},
we also present the MP values of quantum-defects obtained by fitting
Rydberg series calculated with the potential in Eq. \ref{Eq:potential}. We find generally
good agreement for the leading order quantum-defect $\mu^{(0)}_\kappa$, 
estimated in the two methods. Higher-order QD parameters,
$\mu^{(1)}_\kappa$ and $\mu^{(2)}_\kappa$, calculated in the two methods do not agree well. This is 
due to the sensitivity of these parameters to the value of $\mu^{(0)}_\kappa$.
The values for $\mu^{(0)}_\kappa$ obtained by fitting to the MP-calculated $np$ levels agree to four
significant digits with the values, for $\mu^{(0)}_\kappa$, extracted from the threshold phase shifts in
Fig. \ref{Fig-phasesCmp}. 
 
Using the calculated quantum defects, we predict energy levels for the 
lowest few $np$ states. Table~\ref{Tab_pLevels} lists these energies and
compares them with the present MP calculation and with 
a recent MBPT single-double (SD) calculation~\cite{Safronova99}.  
The accuracy of our many-body calculation 
was estimated by exercising upper and lower bounds on $\mu^{(0)}_\kappa$, 
and a consistent determination of $\mu^{(1)}_\kappa$ and $\mu^{(2)}_\kappa$ to fit 
$7p$ and $8p$ energies. MBPT results are in reasonable agreement with the MP 
calculations and SD predictions for these levels.

\subsection{Cross-section}
\begin{figure}
\centerline{\includegraphics[scale=0.65]{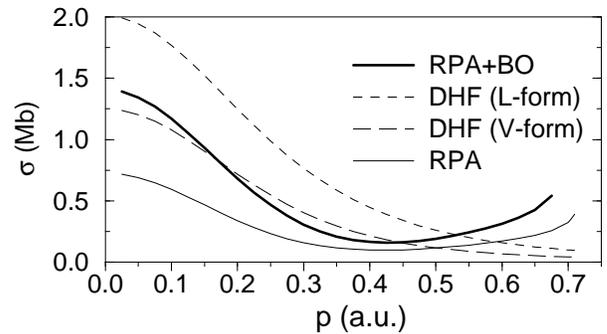}}
\caption{
Total photoionization cross-sections for Fr $7s$ state, 
calculated in various many-body approximations.
\label{fig_xs_all}}
\end{figure}

The total cross-section for photoionization of the valence electron $v$ is 
the sum of partial cross-sections   
\begin{equation}
\sigma = \sum_\kappa \sigma_\kappa =  
\frac{4 \pi^2 \alpha}{3} \omega \, \sum_\kappa   |D_\kappa|^2 \, , 
\end{equation}
where $\omega$ is the photon energy. The dipole transition amplitude for 
an ionization channel $v \rightarrow \epsilon \kappa$ 
is defined as 
\begin{equation}
D_\kappa = 
i^{-l+1} e^{i \bar{\delta}_\kappa} \langle u_{\epsilon \kappa} || {\mathbf r} || u_v \rangle \, ,
\end{equation}
where $u_v$ is the valence wave function and where the continuum wave function $u_{\epsilon \kappa}$ 
is normalized on the energy scale. 
Here we have two ionization channels $7s\rightarrow \epsilon p_{1/2}$, with $\kappa=1$ and
$7s\rightarrow \epsilon p_{3/2}$, with $\kappa=-2$.
The DHF results for the total cross-section are shown with dashed lines
in Fig.~\ref{fig_xs_all}. Since the DHF potential is non-local, the resulting
amplitudes depend on the gauge of the electromagnetic field.  The
difference between length- and velocity-form values is especially noticeable
in the near-threshold region.

Second-order corrections, and the associated all-order sequence
of random-phase approximation (RPA) diagrams, account for the 
shielding of the external field by the core electrons. Explicit expressions for
the second-order MBPT corrections can be found, for
example, in Ref.~\cite{thirdorder}.
Already in second order, the dipole operator with RPA corrections 
reduces at large $r$ to an effective one-particle operator 
\begin{equation}
{\mathbf r}_{\rm eff} =
{\mathbf r} \left(  1 -\frac{\alpha_{\rm d}(\omega)}{r^3}    \right) \, ,
\label{Eq:shield}
\end{equation}
where  $\alpha_{\rm d}(\omega)$ is a {\em dynamic} polarizability of the core.
The first term is associated with the applied electric field 
and the second with the field of the induced dipole moment of the atomic core; 
the valence electron responds to a sum of these two fields. We note that the
induced field may become strong and reverse the direction of the total field.

\begin{figure}
\centerline{\includegraphics[scale=0.45]{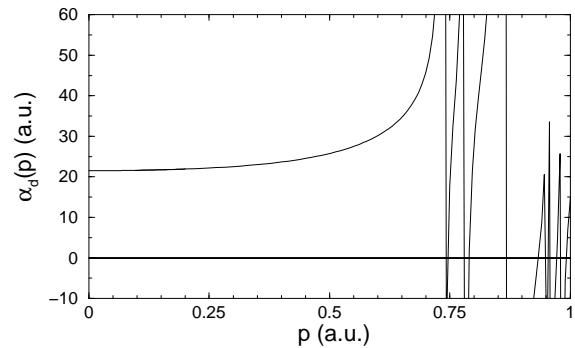}}
\caption{
RRPA dynamic polarizability of Fr$^+$ as a function of photoelectron momentum,
calculated with the $7s$ DHF threshold,  0.13107 a.u..
\label{fig_alpha_p}}
\end{figure}

The RPA cross-section is presented with a thin solid line in Fig.~\ref{fig_xs_all}.
In contrast to DHF amplitudes, the RPA amplitudes are gauge-independent.
Furthermore, we note the sudden upturn
in the  RPA cross-section for the photoelectron momenta $p \approx 0.7$~a.u.\
associated with a $J=1$ core excitation resonance.
To predict the position of this resonance,
we calculate the dynamic polarizability of Fr$^+$ within the framework of relativistic 
RPA, discussed in~\cite{RRPA-pol}. The energy of the first resonance is 
at $\omega_r = 0.4024$~a.u..
Using the DHF value of $7s$ threshold, 0.1311~a.u., 
we expect the first core excitation resonance to appear at $p \approx 0.74$~a.u..
The dynamic polarizability of Fr core $\alpha_{\rm d}(\omega)$ from this 
RPA calculation is plotted as a function of electron momentum $p$ in Fig.~\ref{fig_alpha_p}.

\begin{figure}
\centerline{\includegraphics[scale=0.5]{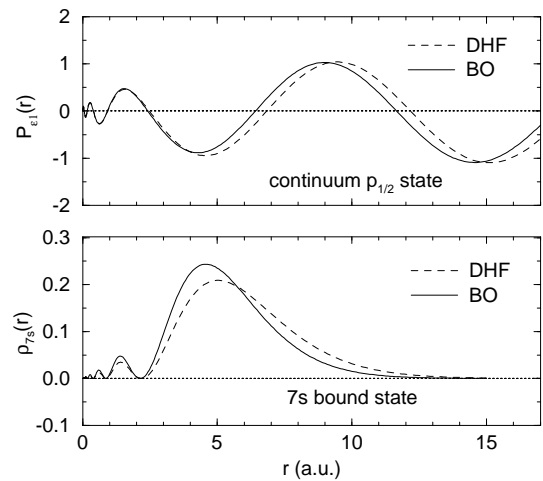}}
\caption{Brueckner (BO) and DHF orbitals. Upper panel:
large component of continuum wave function; Lower panel: radial density. \label{figBO}}
\end{figure}

To account for {\em core-polarization} corrections to the DHF wave function, discussed in the
introduction, 
we evaluate the second-order corrections to the DHF wave functions of the valence electron 
due to the self-energy operator $\Sigma^{(2)}$
\begin{equation}
u_v^{(2)} = \sum_{i \neq v } \frac{\Sigma^{(2)}_{iv}   }{\epsilon_v -\epsilon_i}\,u_i\, .
\end{equation}
The resulting orbital $u_v + u_v^{(2)}$ is the perturbative approximation to 
the valence-state Brueckner orbitals (BO).
Approximate Brueckner orbitals for a continuum state ($\epsilon\, \kappa$) are found
by solving the inhomogeneous Dirac equation
\begin{equation}
 \left( h + V_{N-1} -\epsilon \right)\, w_{\epsilon \kappa} = 
\left( -\pi \sin{\delta_\kappa}
- \Sigma^{(2)}\right)\, u_{\epsilon \kappa}
\end{equation}
normalized on the energy scale, where $\delta_\kappa$ is given in Eq.~(\ref{phase}). 
Brueckner orbitals for the $7s$ valence state and a $p_{1/2}$ continuum state are compared
with unperturbed DHF orbitals in Fig.~\ref{figBO}.

\begin{figure}
\centerline{\includegraphics[scale=0.65]{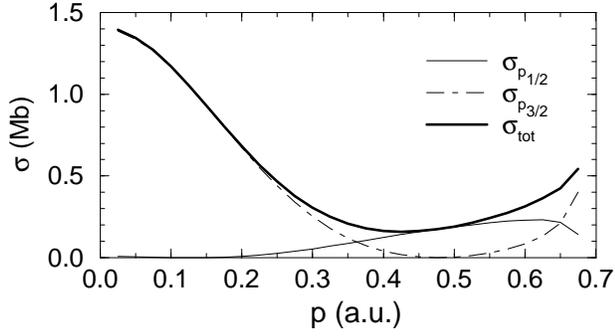}}
\caption{
Partial and total cross-sections for Fr $7s$ 
calculated in the RPA$\oplus$BO many-body approach.}
\label{fig_xs_fin}
\end{figure}

The BO corrections contribute to transition amplitudes starting from third order. 
Together with the RPA corrections, they provide the most important
third-order contributions for bound-bound transitions, as discussed in \cite{thirdorder}.
In the present approach, we modify the conventional RPA scheme by replacing the valence
and continuum wave functions by the approximate Brueckner orbitals 
described above (RPA$\oplus$BO).
Such a modification accounts for the important second- and third-order correlation
corrections and for a subset of fourth-order contributions to transition 
amplitudes. We note that this fourth-order subset brings the photoionization 
parameters in cesium into good agreement with available experimental data;
therefore, we believe that this approach will provide reliable predictions for francium.
The resulting cross section is shown with a heavy solid line in Fig.~\ref{fig_xs_all},
and decomposed into partial cross-sections in Fig.~\ref{fig_xs_fin}.
Calculations using length and velocity forms of transition operator lead to slightly 
different result in the modified RPA$\oplus$BO scheme; we present the 
final result in the length form only. Both photoionization channels exhibit Cooper minima;
$\sigma_{p_{1/2}}$ vanishes at $p\approx 0.1$~a.u. and 
$\sigma_{p_{3/2}}$ vanishes at $p\approx 0.5$~a.u.. Combining the two partial cross sections, 
leads to a broad minimum in the total cross-section
slightly below $p=0.45$~a.u.. 
The total cross section in Fig.~\ref{fig_xs_fin} is not very
sensitive to the positions of Cooper minima in the $p_{1/2}$ and $p_{3/2}$
channels. Conversely, the spin-polarization and angular distribution
measurements, discussed in the following
section, provide information sensitive to fine details of individual transition amplitudes. 
 
\begin{figure}
\centerline{\includegraphics[scale=0.65]{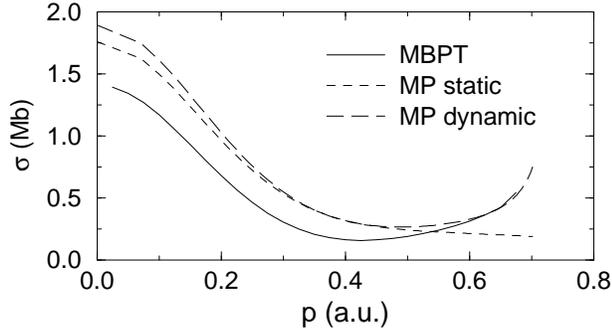}}

\caption{
Comparison of Fr $7s$ photoionization cross sections calculated with
the MBPT and 
the MP methods. 
The curves labeled as "static" and "dynamic" refer, respectively, to the calculations with 
static and dynamic core polarizabilities in Eq.~\ref{Eq:shield}.}
\label{XSecMP}
\end{figure}

Fig.~\ref{XSecMP} examines the total photoionization cross sections for francium, calculated 
in the two method. 
The label "static" refers to the set of MP results with the core static dipole polarizability 
in Eq.~\ref{Eq:shield}.
The shielding of the electron dipole operator is truncated in the MP calculations by introducing a cut-off 
term, similar to the exponential term in the one-electron potential in Eq.~\ref{Eq:potential}.
The threshold cross sections in the $p_{1/2}$ and $p_{3/2}$ channels (not shown here) 
are, respectively,
1.74 and 0.02 Mb. Cooper minima appear in both channels at 
approximately, $p\approx 0.15$ a.u. and $p\approx 0.75$ a.u.
and the maximum cross section in the $p_{1/2}$ channel is $\sigma_{p_{1/2}}({\rm max})\approx 0.2$ Mb. 
The Cooper minimum in the $p_{3/2}$ photoelectron cross section calculated in the MP method with the 
static core dipole polarizability, occurs approximately
where the first core 
resonance in Fig.~\ref{fig_alpha_p} becomes excited. By including the dynamic core polarizability 
$\alpha_d(\omega)$ in the MP calculations, the curve labeled as "dynamic" 
in Fig.~\ref{XSecMP} is obtained. The Cooper minima are moved to lower photoelectron momenta, resulting
in a shallow minimum in the total cross section near 
$p\approx 0.5$ a.u. 

The comparison in Fig. \ref{XSecMP} indicates that the cross sections calculated in the MP method 
are in general larger than the MBPT cross sections. The ``MP dynamic'' and the MBPT cross sections both
rise for the photoelectron momenta $p> 0.5$ a.u. to meet the first core-excited resonance near 
$p\sim 0.75$ a.u.

\subsection{Polarization parameters}
\begin{figure}
\centerline{\includegraphics[scale=0.65]{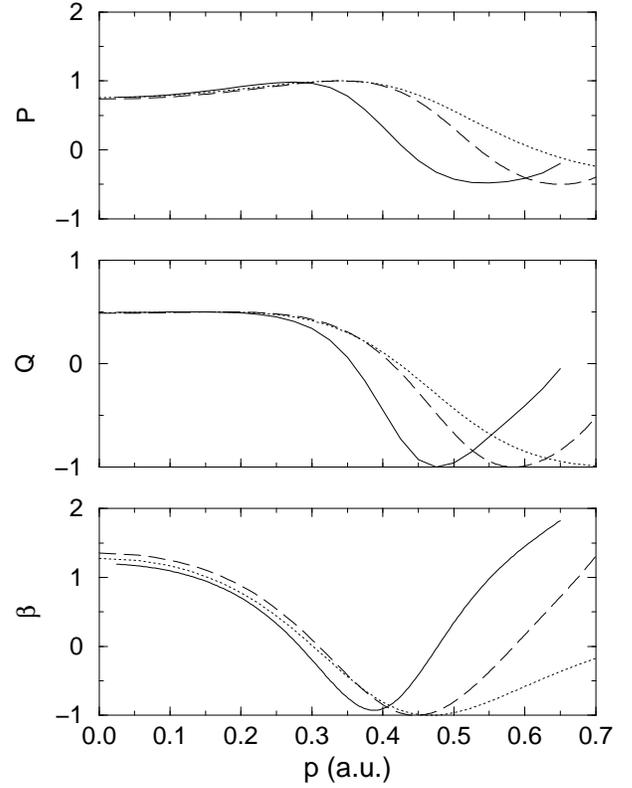}}
\caption{
Fano $P$ and Lubell-Raith $Q$ spin-polarization parameters, and
dipole asymmetry parameter $\beta$,  
calculated in the RPA$\oplus$BO (solid line) and MP approach.
The long-dashed curve  represents the MP results
using the RRPA dynamic polarizability in the transition operator, and
dotted curve --- static polarizability instead.
}
\label{fig_PQb}
\end{figure}

Fano~\cite{Fano69} proposed a measurement of spin polarization $P$ of 
photoelectrons emitted from unpolarized Cs atoms illuminated 
by circularly polarized photons.  The total spin polarization is 
expressed in terms of $p_{1/2}$ and $p_{3/2}$ transition amplitudes
as~\cite{Huang81}
\footnote{ There is a phase difference in the $D_{1/2}D_{3/2}^*$ interference term
 in Eqs.~\ref{EqnP},\ref{Eqnb} and the corresponding equations in Ref.~\protect\cite{Huang81}, 
 caused by
 the unconventional definition of reduced matrix elements used in that work.
} 
\begin{equation}
P = \frac{5\, |D_{3/2}|^2 - 2\, |D_{1/2}|^2 + 4\sqrt{2} \, \Re[D_{1/2} D_{3/2}^*] 
}{6(|D_{3/2}|^2 +  |D_{1/2}|^2) }
 \,.
\label{EqnP}
\end{equation}
The result of our RPA$\oplus$BO calculation of the spin-polarization 
parameter $P$ is presented in Fig.~\ref{fig_PQb}, where it is seen that the
polarization reaches
100\% at momentum $p\approx 0.3$~a.u. The model-potential results for $P$ are  also given in 
Fig.~\ref{fig_PQb} and compare
well with the RPA$\oplus$BO calculation in Fig.~\ref{fig_PQb}. Maximum spin polarization in the MP method 
occurs at $p\approx 0.35$ a.u. 
The calculations with the static and dynamic core polarizabilities in Eq.~\ref{Eq:shield}
are similar and differ only after the maximum is reached.             

Lubell and Raith~\cite{LubellRaith69} measured
a different  spin-polarization parameter $Q$ obtained from photoionization
of polarized Cs atoms by a circularly polarized light.   
In the Lubell-Raith setup, the $p_{3/2}$ channel can be accessed individually,
for example, by photoionization with left-circularly polarized light 
of the $7s$, electron prepared in the $m_s = +\frac{1}{2}$ substate. Combining the partial
cross section $\sigma_{p_{3/2}}$ thereby obtained
with the total cross-section, permits one to deduce the partial cross-section for the $p_{1/2}$
channel.  
The Lubell-Raith parameter $Q$ is defined as the ratio of the difference to the total 
of the photoabsorption intensities for two photon helicities
\begin{equation}
Q = 
\frac{I_{+} - I_{-}}{I_{+} + I_{-}} 
= 
 \frac{ |D_{3/2}|^2 - 2\, |D_{1/2}|^2 }{2(|D_{3/2}|^2 +  |D_{1/2}|^2) }
 \, .
\label{EqnQ}
\end{equation}

The limiting values for the Lubell-Raith parameter are $-1\leq Q\leq \frac{1}{2}$. 
We stress that a measurement of $Q$ or of the (phase-insensitive) parameter $P$,
together with a measurement of the total cross-section permits one to obtain information 
about {\em absolute}
values of transition amplitudes. A further measurement
of the phase-sensitive 
angular-distribution parameter $\beta$~\cite{Huang81},
\begin{equation}
 \beta = 
\frac{|D_{3/2}|^2 - 2\sqrt{2}\Re\left[ D_{1/2} D^*_{3/2} \right] }{|D_{3/2}|^2 + |D_{1/2}|^2} \, ,
\label{Eqnb}
\end{equation}
would permit one to determine the relative phase between the $p_{3/2}$ and $p_{1/2}$
continuum amplitudes and would constitute an essentially  
{\em complete} description of the photoionization process.   
The many-body result for $\beta$ is shown in Fig.~\ref{fig_PQb}.   
The differential cross section is proportional to $1-\frac{1}{2}\beta P_2(\cos{\theta})$.
 
The MP results for $Q$ and the asymmetry parameter $\beta$ are given also 
in Fig.~\ref{fig_PQb}. The 
comparison between MP and MBPT results is generally favorable; the results with the core dynamics
polarizability in Eq.~\ref{Eq:shield} are in better qualitative agreement with the MBPT calculations. We note
that near $p\approx 0.45$ a.u., 
the photoelectron has the propensity to be ionized perpendicular to the 
photon polarization axis and near $p\approx 0.6$ a.u., the photoelectron is preferentially ejected in the   
$j=\frac{1}{2}$ channel, where $Q\rightarrow -1$. A similar situation is evident from MBPT
results at 
$p\approx 0.5$. 
The other limiting value is reached near threshold, 
where $\sigma_{p_{1/2}}\rightarrow 0$.

\section{Conclusion}

We have calculated the photoionization cross sections of the ground-state francium. 
Both many-body and 
model-potential approaches were employed to obtain the cross sections, quantum defects, 
spin-polarization parameters and photoelectron
asymmetry parameter. We find Cooper minima in both $p_{1/2}$ and $p_{3/2}$ channels. 
The comparison between the MBPT and MP results are satisfactory. The Cooper minima predicted in the 
MP calculations are at higher photoelectron energies than those calculated in the MBPT method. The origin of this 
difference can be traced to the shielding of the valence-electron dipole by the core
electrons. The induced dipole  moment 
of the core manifests itself as a dynamic polarizability term.
Upon replacing the static core polarizability with the dynamic
polarizability, 
better quantitative agreement with the MBPT results is observed. 
We predict the energy-dependence of the photoelectron spin-polarization 
and asymmetry parameters which we hope will
stimulate further experimental work in francium.

\section*{Acknowledgments}
The work of AD and WRJ was supported in part by NSF Grant No.\ PHY 99-70666. 
HRS is supported by a grant by NSF to the Institute for Theoretical Atomic and Molecular Physics. 
The authors owe a debt of gratitude to Harvey Gould for describing his proposed
measurement of $\sigma$ and $Q$ for francium.
\newpage



\begin{table}
\caption{Predicted quantum-defect parameters \label{Tab_mu}}
\begin{center}
\begin{tabular}{clll}
 state  &    
\multicolumn{1}{c}{$\mu^{(0)}_\kappa$} & 
\multicolumn{1}{c}{$\mu^{(1)}_\kappa$} & 
\multicolumn{1}{c}{$\mu^{(2)}_\kappa$} \\
\hline
  \multicolumn{4}{c}{MBPT}\\
 $p_{1/2}$  & 4.620(3)  & -0.3(1) & 5.3(7)\\
 $p_{3/2}$  & 4.517(3)  & -0.5(1) & 4.7(8)\\[0.5pc]      
  \multicolumn{4}{c}{Model Potential}\\
 $p_{1/2}$  & 4.605  & -0.79 & 1.61\\
 $p_{3/2}$  & 4.505  & -0.87 & 1.59\\[0.5pc]      
\end{tabular}
\end{center}
\end{table} 

\begin{table}
\caption{Predicted $p$ levels of Fr in cm$^{-1}$. \label{Tab_pLevels}}
\begin{center}
\begin{tabular}{cccc}
 \multicolumn{1}{c}{} &
 \multicolumn{2}{c}{Present Work} \\
 \cline{2-3}
 state  
& QDT+MBPT\tablenotemark[1] 
& MP  
& SD\tablenotemark[2] \protect\cite{Safronova99} \\
\hline
 $9p_{1/2}$   & 5748(3) & 5737   & 5738   \\
 $10p_{1/2}$  & 3800(2) & 3790   & 3795 		      \\
 $11p_{1/2}$  & 2700(2) & 2691   &			      \\
 $12p_{1/2}$  & 2016(1) & 2011   &			      \\[0.5pc]
 $9p_{3/2}$   & 5496(2) & 5487   & 5488 \\
 $10p_{3/2}$  & 3662(2) & 3655   & 3659 		      \\
 $11p_{3/2}$  & 2616(2) & 2610   &			      \\
 $12p_{3/2}$  & 1962(1) & 1958   &			      \\
\end{tabular}
\end{center}
\tablenotetext[1]{The error bars were estimated based on the agreement~\protect \cite{WRJ_Indian97}
of the predicted second-order phase shifts with known quantum defects  
for cesium $p$-states.}
\tablenotetext[2]{Values marked as predicted in Ref.~\cite{Safronova99}.  }
\end{table} 


\begin{references}


\bibitem{exf}              J. E. Simsarian, W. Z. Zhao, L. A. Orozco and G. D. Sprouse,
                          Phys.\ Rev.\ A {\bf 59}, 195 (1999); 
			  G. D. Sprouse, L. A. Orozco, J. E. Simsarian and W. Z. Zhao,
			  Nucl.\ Phys.\ A {\bf 630}, 316C (1998);
			  J. E. Simsarian,  L. A. Orozco, G. D. Sprouse and W. Z. Zhao,
                          Phys.\ Rev.\ A {\bf 57}, 2448 (1998); 
			  W. Z. Zhao, J. E. Simsarian,  L. A. Orozco, W. Shi and G. D. Sprouse,
                          Phys.\ Rev.\ Lett.\ {\bf 78}, 4169 (1998); 
			  G. D. Sprouse, L. A. Orozco, J. E. Simsarian, W. Shi and W. Z. Zhao,
                          Optics Lett.\ {\bf 21}, 1939 (1996). 
             

\bibitem{thf}             A. Derevianko, W. R. Johnson, M. S. Safronova and J. F. Babb,
                          Phys.\ Rev.\ Lett.\ {\bf 82}, 3589 (1999);
			  T. M. R. Byrnes, V. A. Dzuba, V. V. Flambaum and D. W. Murray,
			  Phys.\ Rev.\ A {\bf 59}, 3082 (1999);
			  W. A. Van Wijngaarden and J. Xia,
			  J. Quant.\ Spectrosc.\ \& Radiat.\ Transfer {\bf 61}, 557 (1999);
			  E. Biemont, P. Quinet and V. Van Renterghem,
			  J. Phys.\ B {\bf 31} 5301 (1998);
			  A. Owusu, R. W. Dougherty, G. Gowri, T. P. Das and J. Andriessen;
			  Phys.\ Rev.\ A {\bf 56}, 305 (1997).
			  W. R. Johnson, Z. W. Liu and J. Sapirstein,
			  At.\ Data Nucl.\ Data Tables {\bf 64}, 279 (1996);
			  V. A. Dzuba, V. V. Flambaum, and O. P. Sushkov,
                          Phys.\ Rev.\ A{\bf 51}, 3454 (1995).
                          E. Eliav, U. Kaldor and Y. Ishikawa,
                          Phys.\ Rev.\ A {\bf 50}, 1121 (1994).
			  

\bibitem{Marinescu98}     M. Marinescu, D. Vrinceanu, and H. R. Sadeghpour,
                          Phys.\ Rev.\ A{\bf 58}, R4259 (1998).


\bibitem{HG}              Harvey Gould, private communication.


\bibitem{CK}              J. J. Chang and H. P. Kelly, 
                          Phys.\ Rev.\ A {\bf 5}, 1713 (1972).  	      

\bibitem{OM}              W. Ong and S. T. Manson, 
                          Phys.\ Rev.\ A {\bf 20}, 2364 (1979).

\bibitem{HS}              K.-N. Huang and A. F. Starace, 
                          Phys.\ Rev.\ A {\bf 19}, 2335 (1979).

\bibitem{JS}              W. R. Johnson and G. Soff,             
                          Phys.\ Rev.\ Lett.\ {\bf 50}, 1361 (1983).

\bibitem{FJ}              M. G. J. Fink and W. R. Johnson, 
                          Phys.\ Rev.\ A {\bf 34}, 3754 (1986). 				 


\bibitem{Fano69}          U. Fano, 
                          Phys.\ Rev.\ {\bf 178}, 131 (1969);
			  U. Heinzmann, J. Kessler and J. Lorenz,
			  Z. Phys.\ A {\bf 240}, 42 (1970).
                        

\bibitem{LubellRaith69}   M.\ S.\ Lubell and W.\ Raith, 
                          Phys.\ Rev.\ Lett.\ {\bf 23}, 211 (1969);
			  G. Baum, M. S. Lubell and W. Raith,
			  Phys.\ Rev.\ Lett.\ {\bf 25}, 267 (1970).


\bibitem{Yin}             Y. Y. Yin and D. S. Elliott,
                          Phys.\ Rev.\ A {\bf 46}, 1339 (1992).


\bibitem{WRJ_Indian97}    W. R. Johnson, 
                          Indian J. Phys.\ {\bf 71B}, 263 (1997).


\bibitem{Hameed}          S. Hameed, A. Hertzenberg, and M. G. James,
                          J. Phys.\ B {\bf 1}, 822 (1968);
                          S. Hameed, Phys.\ Rev.\ {\bf 179}, 16 (1969).

\bibitem{Beigman}         I. L. Beigman, L. A. Vainshtein, and V. P. Shevelko,
                          Opt.\ Spectrosc.\ {\bf 28}, 229 (1970).

\bibitem{WD}              J. C. Weisheit and A. Dalgarno,
                          Phys.\ Rev.\ Lett.\ {\bf 27}, 701 (1971).

\bibitem{Weisheit}        J. C. Weisheit, 
                          Phys.\ Rev.\ A{\bf 5}, 1621 (1972).

\bibitem{Norcross}        D. W. Norcross,
                          Phys.\ Rev.\ A {\bf 7}, 606 (1973).


\bibitem{expt}            E. Arnold \etal, 
                          J. Phys.\ B {\bf 22}, L391(1989);
                          J. Phys.\ B {\bf 25}, 3511(1990);
                          J. Bauche \etal, 
                          J. Phys.\ B {\bf 19}, L593(1986);
                          H. T. Duong \etal, 
                          Europhys.\ Lett.\ {\bf 3}, 175(1987);
                          S. V. Andreev, V. S. Letokhov and V. I. Mishin, 
                          \prl {\bf 59}, 1274(1987); 
                          J. Opt.\ Soc.\ Am.\ B {\bf 5}, 2190(1988);
                          S. Liberman \etal, 
                          C. R. Acad.\ Sci.\ Paris {\bf 268B}, 253(1987).

 
\bibitem{Seaton58}         M. J. Seaton,
                           Mon.\ Not.\ R. Astron.\ Soc.\ {\bf 118}, 504 (1958);
                           U. Fano, 
			         Phys.\ Rev.\ A {\bf 2}, 353 (1970).


\bibitem{msd}              M. Marinescu, R. H. Sadeghpour, and A. Dalgarno,
                           \pra {\bf 49}, 982 (1994); 
                           C. H. Greene, \pra {\bf 42}, 1405(1990).



\bibitem{Safronova99}      M. S. Safronova,  W. R. Johnson, and 
                           A. Derevianko (submitted to Phys.\ Rev.\ A, 1999).


\bibitem{Huang81}          K.-N. Huang, W. R. Johnson, and K. T. Cheng,
                           At.\ Data Nucl.\ Data Tables {\bf 26}, 33 (1981).


\bibitem{RRPA-pol}         A. Dalgarno and W. D. Davidson,
                           Adv.\ At.\ Mol.\ Phys.\ {\bf 2}, 1 (1966);
                           D. Kolb, W. R. Johnson, and P. Shorer,
                           Phys.\ Rev.\ A{\bf 26}, 19 (1982);
                           W. R. Johnson, D. Kolb, and K.-N. Huang, 
                           At.\ Data  Nucl.\ Data Tables {\bf 28}, 333 (1983).

\bibitem{thirdorder}       W. R. Johnson, Z. W. Liu, and J. Sapirstein, 
                           At.\ Data Nucl.\ Data Tables {\bf 64}, 280 (1996).


\end{references}
\end{document}